# A spatiotemporal knowledge graph-based method for identifying individual activity locations from mobile phone data

**Jian Li (Corresponding author)**
Associate Professor, Ph.D.
Key Laboratory of Road and Traffic Engineering of the Ministry of Education, Tongji University
College of Transportation & Urban Mobility Institute
4800 Cao'an Road, Shanghai, 201804, China
Email: jianli@tongji.edu.cn

**Tian Gan**
Ph.D. Candidate
Key Laboratory of Road and Traffic Engineering of the Ministry of Education, Tongji University
College of Transportation
4800 Cao'an Road, Shanghai, 201804, China
E-mail: tiangan@tongji.edu.cn

**Weifeng Li**
Associate Researcher
Key Laboratory of Road and Traffic Engineering of the Ministry of Education, Tongji University
College of Transportation
4800 Cao'an Road, Shanghai, 201804, China
Email: liweifeng@tongji.edu.cn

**Yuhang Liu**
Graduate Student
Key Laboratory of Road and Traffic Engineering of the Ministry of Education, Tongji University
College of Transportation
4800 Cao'an Road, Shanghai, 201804, China
E-mail: jeremylaw@tongji.edu.cn

# ABSTRACT

In recent years, mobile phone data has been widely used for human mobility analytics. Identifying individual activity locations is the fundamental step for mobile phone data processing. Current methods typically aggregate spatially adjacent location records over multiple days to identify activity locations. However, only considering spatial relationships while overlooking temporal ones may lead to inaccurate activity location identification, and also affect activity pattern analysis. In this study, we propose a spatiotemporal knowledge graph-based (STKG) method for identifying activity locations from mobile phone data. An STKG is designed and constructed to describe individual mobility characteristics. The spatial and temporal relationships of individual stays are inferred and transformed into a spatiotemporal graph. The modularity-optimization community detection algorithm is applied to identify stays with dense spatiotemporal relationships, which are considering as activity locations. A case study in Shanghai was conducted to verify the performance of the proposed method. The results show that compared with two baseline methods, the STKG-based method can limit an additional 45% of activity locations with the longest daytime stay within a reasonable spatial range; In addition, the STKG-based method exhibit lower variance in the start and end times of activities across different days, performing approximately 10% to 20% better than the two baseline methods. Moreover, the STKG-based method effectively distinguishes between locations that are geographically close but exhibit different temporal patterns. These findings demonstrate the effectiveness of STKG-based method in enhancing both spatial precision and temporal consistency.

**Keywords:** Human mobility, Mobile phone data, activity location, Spatiotemporal knowledge graph, Community detection

# 1 INTRODUCTION

Human mobility, which refers to the movement of individuals between geographical locations, is a topic of interest across multiple disciplines (Barbosa et al., 2018). In the past decade, mobile phone data with high spatiotemporal coverage has offered great opportunities for human mobility analysis (Abbiasov et al., 2024; Bonnetain et al., 2021; Dong et al., 2024; Gonzalez et al., 2008; Schläpfer et al., 2021; Xu et al., 2018). Compared with household travel survey data which includes individual's complete information of activities and travels, mobile phone data only generates records when a user connects to certain cell tower. The temporal resolution of mobile phone data is typically sparse, and its spatial resolution is also lower than that of household travel survey data. In addition, the location of mobile phone records at the same activity location may deviate due to the distribution of cell towers, which is defined as the locational uncertainty of mobile phone sightings data (Wang and Chen, 2018). As a result, the mobile phone data needs to be processed first to obtain the information of activity locations, trip purposes, travel modes and routes. Identification of activity locations is the first step in extracting human mobility related information from mobile phone data. The accuracy of the method is crucial for subsequent information extraction and analysis.

Current studies typically identify activity locations based on spatial correlation of mobile phone records. One simple and straightforward method is to predefined a spatial range (Jiang et al., 2013, 2017; Alexander et al., 2015; Widhalm et al., 2015; Wang et al., 2018), e.g. 500m*500m grid, and records within certain grid are considered to belong to the same activity location (Qian et al., 2021; Xu et al., 2018). However, the actual boundaries of activity locations may vary, the predefined spatial threshold may cause accuracy issue. Some studies use spatial clustering methods to identify activity locations by adaptively aggregating spatial adjacent records (Lv et al., 2016; Bachir et al., 2019; Huang et al., 2023). Although these methods, e.g. DBSCAN, don't need to predetermine a spatial range, the parameters of search radius and minimum points still need to be well fitted. Moreover, the above two methods only identify activity locations based on the spatial similarity, they may not be able to identify activities which are spatially adjacent but temporally distinct. For example, a resident whose workplace and residence are close to each other, the current method might merge the work and home as one activity location, or mistake part of the work-related activity location to the home-related one, or vice versa. In fact, the above example is not uncommon in residents' daily activities, previous studies show that a considerable number of activities occur around residence or workplace locations (Golledge and Stimson, 1997; Hasanzadeh, 2019; Perchoux et al., 2013).

It is a promising approach for identifying activity locations by considering both spatial and temporal proximity of mobile phone records, but it also introduces challenges (Ansari et al., 2020). The spatiotemporal clustering methods, e.g. ST-DBSCAN (Birant and Kut, 2007), has been used for mobile phone data processing (Yao et al., 2022). However, because ST-DBSCAN is an extension of DBSCAN, its performance still heavily relies on user-specified parameters (Deng et al., 2013). To eliminate the above shortcomings, a feasible approach is to transform mobile phone records into a graph-based data structure, which can intuitively capture the complex spatiotemporal relationships within a multidimensional space (Ferreira et al., 2020). To identify activity locations, parameter-free graph partitioning algorithms can be used to find subgraphs with strong spatiotemporal relationships. The key issue is how to construct a graph that simultaneously represent both spatial and temporal relationships of mobile phone records. Thanks to the emerging knowledge graph (KG), which provides an efficient and easily operable graph-based data structure for computing large interconnected data sets (Del Mondo et al., 2010). A knowledge graph uses heads, tails, and relations to describe human knowledge, enabling machines to comprehend and reason more efficiently (Hogan et al., 2021). By representing spatiotemporal entities and their connections as a spatiotemporal knowledge graph (STKG), it can potentially enhance the performance of mobile phone data processing (Liu et al., 2022; Wang et al., 2021).

This study proposes an STKG-based method for identifying individual activity locations from mobile phone data. An STKG is designed and constructed to describe individual mobility characteristics. The spatial and temporal relationships of individual stays are inferred and transformed into a spatiotemporal graph. The modularity-optimization community detection algorithm is applied to identify stays with dense spatiotemporal relationships, which are considering as activity locations. A case study in Shanghai was conducted to verify the performance of the advantages of the proposed method.

The main contributions can be summarized as follows.

First, we proposed a method for constructing and expanding an STKG based on mobile phone data. By conceptualizing the STKG through the understanding of individual mobility pattern, an STKG for human mobility is constructed. Spatial and temporal relationships are inferred within the STKG, providing concise and accurate spatiotemporal knowledge for identifying activity locations.

Second, we proposed an STKG-based method for identifying activity locations. Activities occurring at the same activity location are assumed to be not only spatially adjacent but also temporally co-occurring. By constructing a spatiotemporal graph through the STKG, nodes with strong spatiotemporal correlations are identified and mapped as activity locations.

The remainder of the paper is structured as follows. Section 2 reviews the activity identification method based on mobile phone data and the application of STKG. Section 3 describes the data and the preprocessing procedures used in this study. Section 4 explains the proposed STKG-based method. Section 5 compares the STKG-based method with spatial-constraint-based and non-spatial-constraint methods. Finally, Section 6 summarizes the study and discusses its limitations.

## 2 RELATED WORKS

### 2.1 Activity Location Identification Based on Mobile Phone Data

In current studies, activity location identification methods based on mobile phone data can be classified into two categories: the spatial-constraint based method and non-spatial-constraint method.

The spatial-constraint based method usually divide the research area into several sub-areas, and mobile phone records within certain sub-areas are considered to belong to the same activity location. The most straightforward way is to divide the study area into grids (Qian et al., 2021; Xu et al., 2018). A more general approach is to predefine spatial range of sub-areas, and then identify specific activity location within this range (Jiang et al., 2013, 2017; Alexander et al., 2015; Widhalm et al., 2015; Wang et al., 2018). The spatial-constraint based method draws on the GPS data processing technique (Jiang et al., 2013). However, the spatial accuracy of mobile phone data is lower than that of GPS data, and the boundaries of activity locations are not unique. Consequently, there is no universal predefined spatial threshold that applies across different regions and datasets.

The non-spatial-constraint method or DBSCAN-based method does not impose spatial constraints (Lv et al., 2016; Bachir et al., 2019; Huang et al., 2023). It can handle variations in activity location boundaries. Clusters with irregular shapes can be identified by two parameters: the epsilon and minimum number of samples. However, it is also tricky for parameters setting, particularly in cases where data points are evenly distributed and sparse regions are minimal (Aragones and Ferrer, 2024). In such cases, the sensitivity to parameter settings may lead to difficulties in finding an appropriate clustering structure (Hou et al., 2016). If the activity points that are far apart are classified into one cluster, the range of identified activity point may be too large (Wang and Chen, 2018).

### 2.2 Spatiotemporal Data Clustering

Mobile phone data is a typical type of spatiotemporal data. Activity location identification based on mobile phone data needs to consider both spatial and temporal characteristics. Spatiotemporal clustering algorithm are generally extensions of density-based and other classical clustering algorithms (Kisilevich et al., 2010). In human mobility data processing, spatiotemporal clustering algorithms have already been used, such as identification of trip ends (Fu et al., 2016, 2016; Ozer et al., 2024; Wang et al., 2019). Most of the algorithms used in these studies integrate spatiotemporal attributes to expand DBSCAN (Yao et al., 2022), e.g., ST-DBSCN (Birant and Kut, 2007). The sensitivity of these algorithms to parameter settings may influence the clustering outcomes, potentially affecting the accuracy of activity location identification.

To avoid shortcomings caused by parameter settings, one approach is to transform the task from spatiotemporal data clustering to spatiotemporal graph partitioning. This method has two advantages. First, the graph-based representation can intuitively depict complex spatiotemporal relationships and achieve dimensionality reduction (Ferreira et al., 2020). Second, the popular parameter-free graph partitioning or community detection algorithms can be applied on the spatiotemporal graph to achieve spatiotemporal clustering, such as modularity optimization algorithms (Blondel et al., 2008; Newman, 2006). The divided subgraphs or communities composed of nodes with strong spatiotemporal relationships are equivalent to clusters with high spatiotemporal similarity.

With the emergence of knowledge graph (KG), the representation, storage, and management of knowledge in graph form has become more efficient (Hogan et al., 2021). KG is easier to map spatiotemporal data into a spatiotemporal graph. In recent years, the application of knowledge graphs has expanded into transportation field, such as the use of public transportation KG for epidemic contact tracing (Chen et al., 2022) and predicting metro passenger flows (Zeng and Tang, 2023). Current studies primarily focus on modeling specific transportation systems. Some have incorporated both spatial and temporal dimensions within human mobility, leading to the development of more comprehensive spatiotemporal knowledge graph (STKG) for urban computing (Liu et al., 2022), such as for mobility prediction (Wang et al., 2021).

## 2.3 Summary

In current studies, activity location identification based on mobile phone data focus on the spatial dimension. These methods may incorrectly merge activity locations that are temporally distinct but spatially close, or inaccurately separate activity locations. Incorporating both spatial and temporal dimensions in activity location identification is essentially a spatiotemporal clustering task. The emergence of knowledge graph (KG) provides a graph-based data structure for organizing spatiotemporal data and facilitates the generation of spatiotemporal graphs. Therefore, this study aims to manage extensive mobile phone records within a spatiotemporal knowledge graph (STKG). By leveraging the STKG, spatiotemporal relationships are inferred and used to identify activity locations by using graph portioning algorithm. The goal is to identify fine-grained spatial details and meaningful activity locations through the STKG.

## 3 PRELIMINARY

Mobile phone sightings data was used in this study. The data refer to records when network events occur, e.g. user's mobile device is connected to certain tower. (Chen et al., 2016; Huang et al., 2019). The raw mobile phone records can be seen in **Table 1**, including the user ID, the timestamp when the record was generated, and the latitude and longitude coordinates. In China, the location information in sightings data refers to the base transceiver station (BTS) to which users' mobile devices are connected.

**Table 1 Example of raw mobile phone records**

| Field | Type | Example |
|---|---|---|
| Uid | String | 9222173994963118848 |
| Timestamp | Timestamp | 2019/6/1 5:55:55 |
| Longitude | Double | 121.2685502 |
| Latitude | Double | 31.34786568 |

The data processing includes filtering out oscillations in the raw data and smoothing mobile phone traces. First, to address the oscillation problem, the displacement and speed between consecutive records, as well as the ratio of the previous to subsequent displacements, were calculated. A threshold-based method was used to filter out oscillation sequences. Next, based on the time difference between consecutive records, the duration for which an individual's mobile phone remains connected to a BTS is calculated. Using this duration as a weight, the weighted average position point corresponding to each 10-minute time slot was computed as the smoothed traces.

To facilitate the storage of raw mobile phone records in the spatiotemporal knowledge graph, the preprocessed individual traces are represented using time slots and grids as basic analysis units in the temporal and spatial dimensions. The day is first divided into 144 time slots, each lasting 10 minutes. Timestamps from the smoothed traces are matched with these time slots. Similarly, the study area is divided into a series of equally-sized grid cells. This allows us to map traces to grids based on their longitude and latitude, using the grids to represent the spatial locations. As a result, the traces are reconstructed with 10 minutes as the basic time unit and grid cells with a length of 500 meters as the basic spatial unit. A trace point can be denoted as $TP_i(G_{(r,c)}, TS_k)$, where $G_{(r,c)}$ represents the grid where the trace is located, $r$ and $c$ represent the row and column indices of grid, and $TS_k$ represents the time slot when the trace occurred.

## 4 METHODS

### 4.1 Framework

As shown in **Figure 1**, the spatiotemporal knowledge graph (STKG) based activity location identification method includes two steps: (1) Conceptualizing and constructing an STKG to store mobility-related spatiotemporal information from large-scale mobile phone data. Then, inferring spatial adjacency and temporal co-occurrence relationships between individual stays to expand the STKG. (2) Representing the spatiotemporal relationship as a spatiotemporal graph. Community detection algorithm is used to divide the graph into densely connected subgraphs, which are then mapped onto geographical locations to form activity locations.

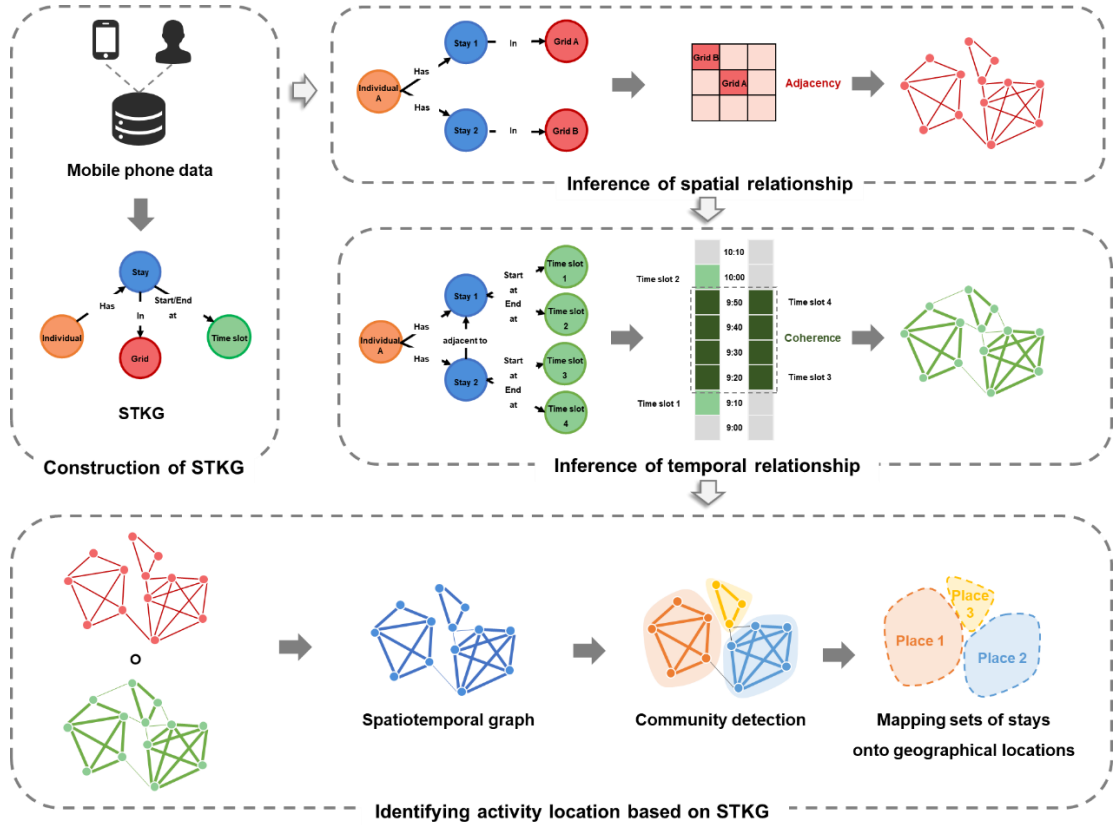

**Figure 1 Framework of STKG-based activity location identification method**

## 4.2 Construction and Expansion of STKG

*4.2.1 Construction of STKG*

Resource Description Framework (RDF) is used to uniformly describe knowledge related to human mobility. Each piece of knowledge is represented as a triple in the format subject-predicate-object. As the basic unit of knowledge graph, the subjects and objects are considered as nodes, while the predicates are treated as edges. A triple represents a relationship through classes and properties. Therefore, before STKG construction, it is essential to define its elements and conceptual model, which depend on the type of knowledge that needs to be stored in the STKG.

**Figure 2** demonstrates the process from individual spatiotemporal path to the STKG of human mobility. An individual's sequence of stays and movements can be represented as a spatiotemporal path that shows their presence in different grid locations at various time slots. In the STKG, the knowledge of what an individual is doing (stay), at what time (time slot), and in what location (grid) can be described using triples. These triples are then stored in a graph format, as illustrated on the right side of **Figure 2**.

Based on the conceptual model, a widely used approach is used to process traces into stays for constructing the STKG. A time threshold is used to determine whether a trace recorded at a specific location represents a stay or a pass-by. Given a sequence of traces for an individual $\{TP_1, TP_2 ... TP_n\}$, for trace $TP_i$ with a time slot of $TS_i$ and a grid unit of $G_{(r1,c1)}$, and $TP_{i+1}$ with a time slot of $TS_{i+1}$ and a grid unit of $G_{(r2,c2)}$, the start time of this user's stay in $G_{(r1,c1)}$, is $TS_i$, and the duration the individual remains in $G_{(r1,c1)}$ is $TS_{i+1} - TS_i$. In this way, the trace is considered a stay if the duration exceeds 10 minutes. Otherwise, it is classified as a pass-by and will not be stored in the STKG.

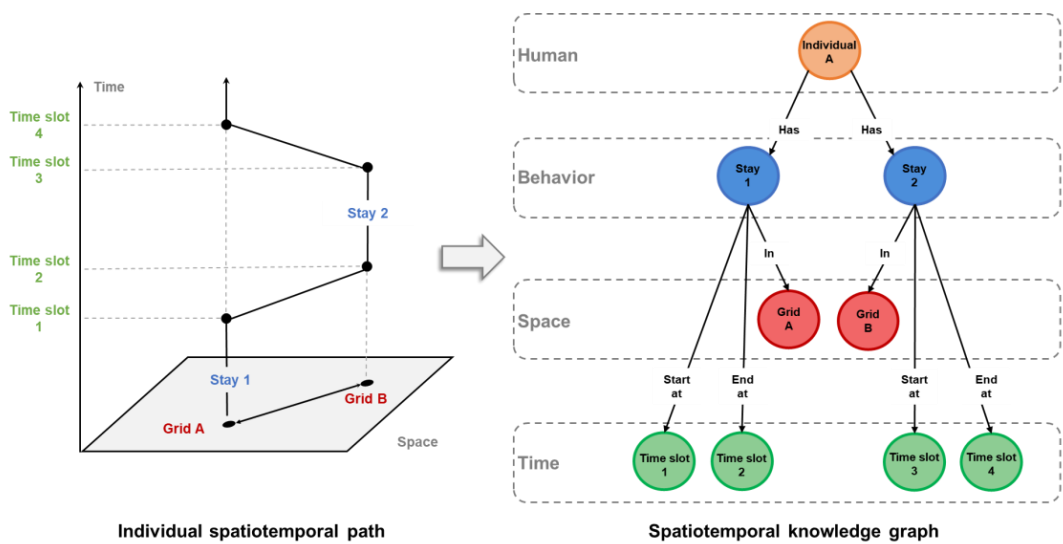

**Figure 2 Illustration of converting a spatiotemporal path to STKG**

*4.2.2 Inference of Spatial Relationship*

The spatial relationships between stays are further inferred based on the STKG. In the STKG, each location is represented by a grid, and spatial relationships are inferred based on these grids. In this grid-based spatial system, the calculation of contiguity weights is based on the definition that two spatial objects are contiguous if they share a common border of non-zero length. Specifically, the more inclusive measure of queen contiguity is used to measure spatial relationships among grid units. Queen contiguity is defined as two spatial objects being contiguous if they share one or more vertices. Therefore, each grid can have up to eight neighbors (excluding itself), including two in the same column, two in the same row, and four on the diagonals.

Let the 2D grid array be $G$, where each cell is uniquely determined by its row and column indices $(r, c)$. For a grid cell $G_{(r,c)}$, its queen contiguity neighbors can be represented as **Equation 1**:

$$N_{(r,c)} = \{(r + a, c + b) \mid a, b \in \{-1, 0, 1\}, (a, b) \neq (0, 0)\} \tag{1}$$

Queen contiguity neighbors all neighbors in the same row, column, and along the diagonals. Since the spatial geographical location of a grid can be uniquely determined through the row and column indices of the grid array, the spatial neighbors can be directly calculated through these indices.

As shown in **Figure 3**, after measuring the spatial relationships between grids, spatial relationships between stays can be inferred through the associations between stays and grids. For an individual who has $n$ stays, the spatial weight between stays can be represented as a $n \times n$ matrix. Elements with a value of 1 indicate that the stays are adjacent to each other, otherwise, the value is 0. Based on the constructed spatial relationships, we further add edges for stays whose grids are not neighboring. If there exists a path connecting these stays in the graph, a spatial relationship is defined between them, resulting in the final spatial relationships.

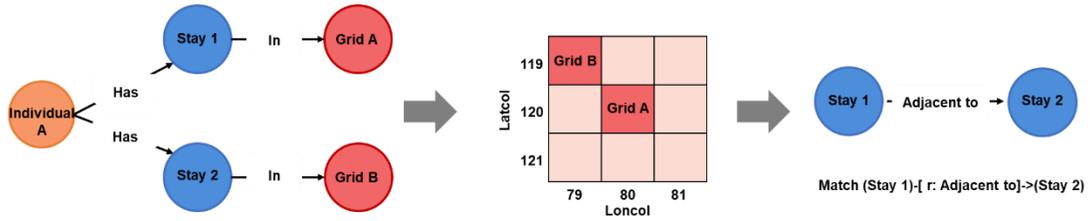

**Figure 3 Example of spatial relationship reference**

*4.2.3 Inference of Temporal Relationship*

The co-occurrence of two stays is defined as both stays occurring within the same period. The longer the intersection between the time periods of the two stays, the higher their co-occurrence degree. As depicted in **Figure 4**, an individual has two stays, stay1 and stay2. Considering 10-minute time slots, the intersection between stay1 and stay2 includes four time slots: 9:20, 9:30, 9:40, and 9:50. This part represents the co-occurrence between the two stays.

Let the time slots of stay1 and stay2 be represented by binary vectors $\boldsymbol{A}$ and $\boldsymbol{B}$, where each vector has a length of 144, corresponding to the 144 time slots (10 minutes each) in a day. A value of 1 in vector $\boldsymbol{A}$ at index $i$ represents that stay1 occupies the $i$-th time slot, and similarly for $\boldsymbol{B}$. To measure the temporal co-occurrence relationship between stay1 and stay2, the cosine similarity between the two vectors is used. This is calculated as **Equation 2**:

$$CS(A,B) = \| A \| \| B \| A \cdot B \tag{2}$$

where $A \cdot B$ represents the dot product of vectors $A$ and $B$, and $\| A \|$ and $\| B \|$ represent the Euclidean norms of the vectors. Since the number of connections to time slots is non-negative, the co-occurrence degree $CS(A,B)$ ranges from 0 to 1, where 0 indicates no overlap and 1 indicates perfect overlap. As shown in **Figure 4**, For an individual, for any pair of spatially connected stays, the temporal co-occurrence degree between them is calculated as **Equation 2**. A temporal relationship is established between the two nodes, with the relationship attribute assigned as the temporal co-occurrence degree, where the maximum value of these elements is 1, representing a perfect overlap in time.

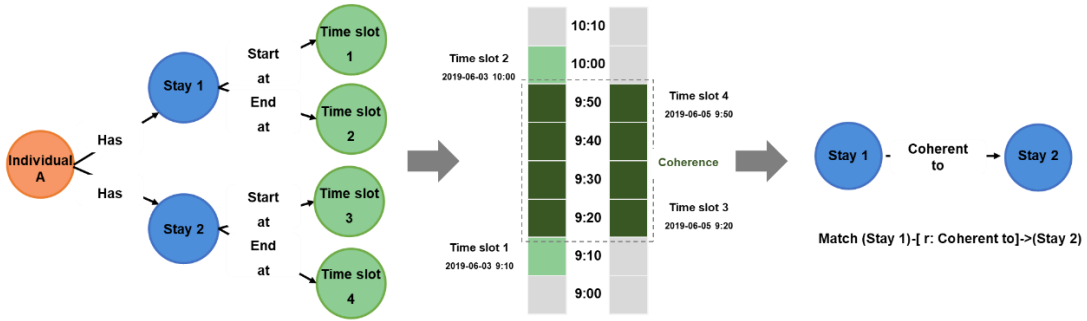

**Figure 4 Example of temporal relationship reference**

**4.3 Identifying Activity Location Based on STKG**

As shown in **Figure 5,** our focus is on dividing each individual stay-based spatiotemporal graph into subgraphs, where each subgraph contains homogeneous stays. These sets of stays form individual activity locations when mapped onto geographical space. For an individual, the graph that depict temporal and spatial relationships is represented as a single graph using the Hadamard product, denoted as $STG = SG \circ TG$, where $SG$ represents the spatial relationship graph, and $TG$ represents the temporal relationship graph. $STG$ is a weighted undirected graph, where each node represents a stay, the edges represent spatiotemporal connections between nodes, and the edge weights indicate the temporal co-occurrence degree between the stays.

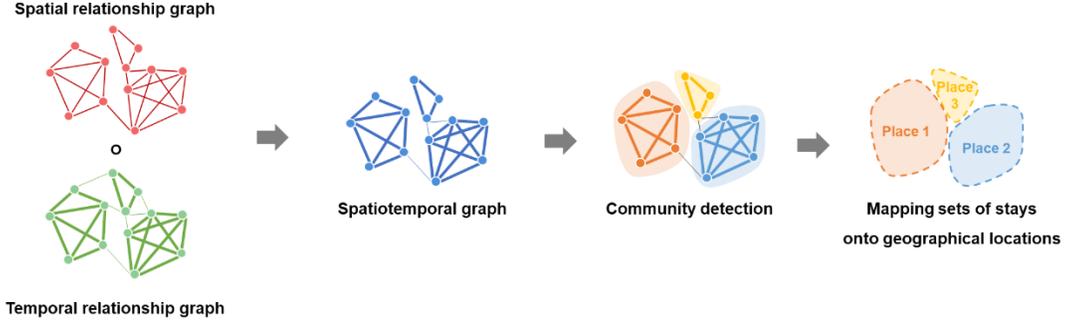

**Figure 5 Illustration of activity locations identification based on spatiotmeporal graph**

Communities in a network are groups of nodes with significantly stronger connections among nodes within the same community compared to those between nodes in different communities (Lancichinetti and Fortunato, 2009). Nodes belonging to the same community are more likely to have similar functions or properties. Therefore, the community detection algorithm is used to divide $STG$ into subgraphs where nodes with dense spatiotemporal relationship.

Fast Unfolding algorithm (Blondel et al., 2008) is used to detecting communities in this study. This well-known optimization algorithm (Su et al., 2022) is recognized for its low computational complexity and fast computation speed (Lancichinetti and Fortunato, 2009). Additionally, it is well-suited for identifying non-overlapping communities in undirected graphs, meaning that each stay corresponds to a single type of activity.

Modularity $Q$ is a key metric for evaluating the quality of community partitioning. Modularity measures the difference between the actual density of edges within communities and the expected density if edges were distributed randomly while preserving the degree distribution. Modularity $Q$ is defined as **Equation 3**:

$$Q = \frac{1}{2m} \sum_{i,j} \left[ A_{ij} - \frac{k_i k_j}{2m} \right] \delta(C_i, C_j)$$

$$m = \frac{\sum_{ij} A_{ij}}{2} \quad (3)$$

$$k_i = \sum_j A_{ij}$$

where $A_{ij}$ is the weight of the edge between nodes $i$ and $j$, $k_i$ is the strength of node $i$. $\delta(C_i, C_j)$ is the Kronecker delta, equal to 1 if nodes $i$ and $j$ belong to the same community ($C_i = C_j$) and 0 otherwise. $C_i$ is the community assignment of node $i$.

In phase 1 of Fast Unfolding algorithm, each node $i$ is initially assigned to its own community. The algorithm evaluates the change in modularity $\Delta Q$ when moving node $i$ to the community $C_j$ of one of its neighboring nodes $j$. The modularity gain $\Delta Q$ is calculated as **Equation 4**:

$$\Delta Q = \left[\frac{W_{in} + 2k_i^{in}}{2m} - \left(\frac{W_{tot} + 2k_i}{2m}\right)^2\right] - \left[\frac{W_{in}}{2m} - \left(\frac{W_{tot}}{2m}\right)^2 - \left(\frac{k_i}{2m}\right)^2\right] \tag{4}$$

where $W_{in}$ is the total weight of the edges inside community $C_j$, $W_{tot}$ is the total weight of the edges connected to nodes in community $C_j$, $k_i$ is the degree (total weight of edges) of node $i$, $k_i^{in}$ is the sum of the weights of edges from node $i$ to nodes in community $C_j$. The algorithm selects the neighboring community $C_j$ that provides the largest positive gain $\Delta Q$. If no positive gain is found, node iii remains in its current community. This process is repeated for all nodes until no further improvement in modularity is possible.

In phase 2 of Fast Unfolding algorithm, the communities obtained in phase 1 are treated as super-nodes in a new network. In this new network, each super-node represents an entire community from phase 1. Then, the weight of edges between the super-nodes is the sum of the weights of the edges between the corresponding communities in the original network. The algorithm then applies phase 1 to this new network of super-nodes. Phases 1 and 2 are repeated iteratively until the modularity $Q$ no longer improves, and a final stable division of the network into communities is achieved.

The community detection helps identify meaningful communities within the graph, grouping stays that exhibit strong spatiotemporal correlations into clusters. The weight, which represents the temporal co-occurrence between stays that are spatially adjacent. By mapping stays that belong to the same community onto geographical locations, one can derive individual activity locations.

## 5 EXPERIMENT DESIGN

### 5.1 Mobile phone dataset

The mobile phone dataset was collected in June 2019 by one of China's largest mobile service providers. The dataset includes a total of 11,734,408 users. The distribution of observed days for users is depicted in **Figure 6 (a)**, where over 16% of users generated data for only one day, and the average observed days are approximately 12 days. As demonstrated in **Figure 6 (b)**, nearly 45% of the daily records of users exceed 60 entries. From the number of records in each hour of the day, as shown in **Figure 6 (c)**, it is evident that daytime hours, due to frequent social and economic activities, result in more mobile phone usage and hence more data generation, whereas nighttime hours show the opposite trend. To validate the effectiveness of the proposed method, experiments were conducted using a sample of 10,000 users randomly selected from the complete mobile dataset.

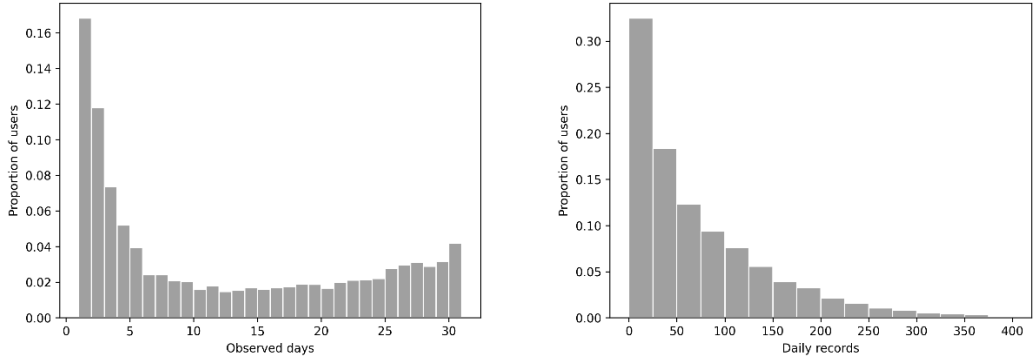

(1) Distribution of observed days of users  (2) Distribution of daily records of users

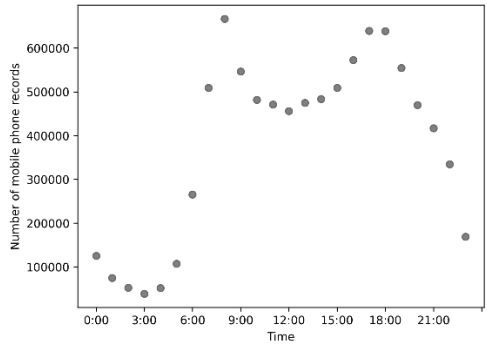

(3) Distribution of records in each hour of the day

**Figure 6 Description of mobile phone dataset**

### 5.2 Benchmarks and parameters

The two benchmarks are briefly summarized as follows.

Spatial-constraint-based method (Jiang et al., 2013). This approach initially processes mobile phone traces into stays using a distance threshold $d$ and a time threshold of 10 minutes (traces with durations not exceeding 10 minutes are considered pass-bys). Subsequently, the stays are sampled using grids with side lengths equal to $d$ representing stays falling within those grid cells. Following this, aggregation is performed based on grid units. In each iteration, the unmarked grid cell containing the maximum number of stays is chosen as the center grid. It is then merged with adjacent grid cells to form a cluster. According to queen contiguity, a grid cell can have a maximum of 8 neighbors, implying that the resulting cluster area is at most $3d \times 3d$. Since in our method the grid size is set to 500 meters per side, $d$ is 500 meters.

Non-spatial-constraint method (Huang et al., 2023). After dressing oscillation issues, the preprocessed traces were directly clustered into clusters using DBSCAN instead of distinguishing between stays and pass-bys. Subsequently, within each cluster, traces with continuous time were divided into stays and pass-bys based on a 10-minute time threshold. For the two parameters in the DBSCAN algorithm, Epsilon ($\varepsilon$) and Minimum Points (MinPts), we first followed Huang et al., 2023 and set MinPts to 1, meaning that as long as a single stay exists within an $\varepsilon$-radius neighborhood, it is considered density-reachable. Furthermore, similar to the STKG-based method and spatial-constraint-based method, the $\varepsilon$-radius neighborhood also represents a similar basic spatial unit concept. Thus, to approximate the grid area based on queen contiguity, we set $\varepsilon$ to $2d$, which is 1000 meters.

## 6 RESULT ANALYSIS

### 6.1 Analysis of Spatial Dimension

The radius of cluster is used to evaluate the activity location identification methods from the spatial perspective. In the spatial dimension, a reasonable activity location is expected to limited to a certain range. This ensures that while resolving the locational uncertainty of the same type of activity on different days, it does not mistakenly mix additional stays and pass-bys within the same spatial area. Therefore, the radius of cluster refers to the distance from the centroid of all stay locations in a cluster to the furthest stay location. Theoretically, the spatial-constraint-based method's radius of cluster does not exceed $3d$, which is 1500 meters. For the non-spatial-constraint method and the STKG-based method, the radius of cluster does not have a fixed upper limit.

As depicted in **Figure 7 (a)**, we first calculated the radius of cluster for all activity locations. Unsurprisingly, in the results identified by the spatial-constraint-based method, the radius of cluster for all activity locations are less than 1500 meters, and nearly all are within 1000 meters. Next, using 1000 meters as the threshold, we calculated the proportion of activity locations with a radius of cluster less than this threshold for the STKG-based method and the non-spatial-constraint method, which are 0.83 and 0.93, respectively. This indicates that, although both methods do not include spatial constraints, the STKG-based method, by considering temporal correlation, achieves more accurate identification for 10% of the activity locations.

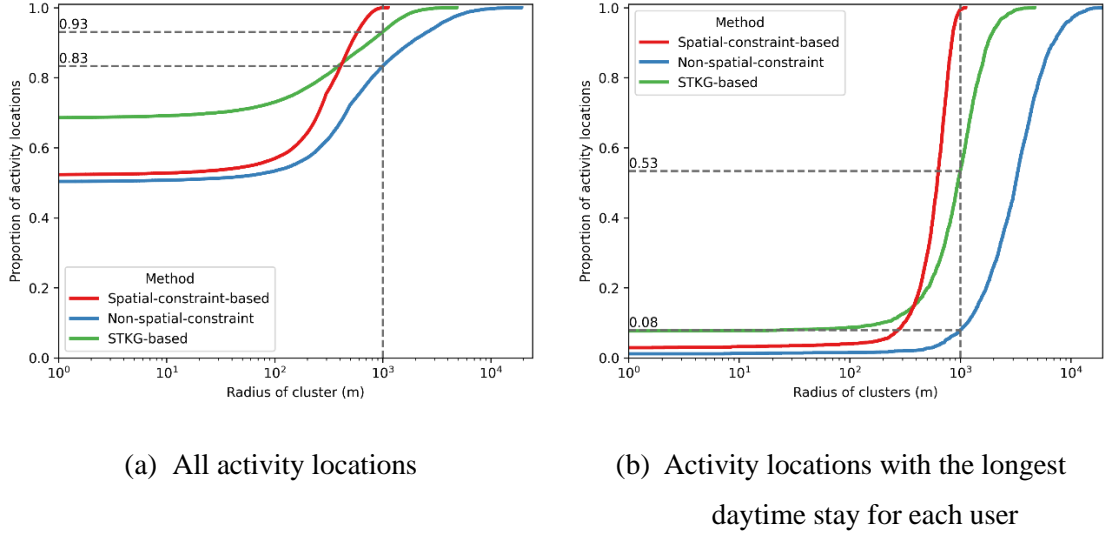

(a) All activity locations  (b) Activity locations with the longest daytime stay for each user

**Figure 7 Result about radius of activity locations**

We further calculated the radius of cluster for activity location with the longest daytime stay for each mobile phone user, as shown in **Figure 7 (b)**. This is because activity locations include many infrequent activity locations, but the identification of frequent activity locations (such as residences or workplaces) are likely the key issue when addressing locational uncertainty in mobile phone data. Users have numerous location records at these activity locations, which increases the likelihood of uncertainty. Specifically, Daytime is defined as from 6 a.m. to 6 p.m. The dates area limited to weekdays, indicating that these activities likely correspond to work activities. The result shows that the STKG-based method has a more significant advantage over the non-spatial-constraint method. While the non-spatial-constraint method may require adjusting input parameters to increase the proportion of activity locations with a radius within 1000 meters, the STKG-based method, with the same parameter settings, can limit an additional 45% of daytime hotspots within a reasonable range.

**6.2 Analysis of Temporal Dimension**

Two indicators are used to evaluate the activity location identification methods from the temporal perspective.

One indicator is the variance of the start and end times of activities occurring at the same activity location on different days. Activities are composed of temporally continuous stays at the same activity location. Misidentification of activity locations can lead to prematurely ending activities or extending activities that should have ended. Therefore, a reasonable activity location is expected to show limited variance in the start and end times of activities, especially for frequent activities, which are likely to follow a more regular pattern.

Another indicator is to measure the number of days on which work activities can be observed to occur. Misidentification of activity locations can lead to activities that should belong to the same location being categorized into different activity locations. Therefore, on weekdays when work activities should occur for users in the working group, misidentification of activity locations may result in our inability to observe the corresponding work activities.

As shown in **Figure 8**, we calculated the variance of the start and end times of activities at activity locations with the longest daytime stay, which may represent the user's workplace. Regardless of whether it is the activity start time or end time, the STKG-based method is more effective in maintaining the variance within a low range, approximately 10% to 20% better than the baseline methods. This is because both the spatial-constraint-based method and the non-spatial-constraint method do not consider the temporal dimension in activity location identification, which can result in activities being incorrectly interrupted or extended.

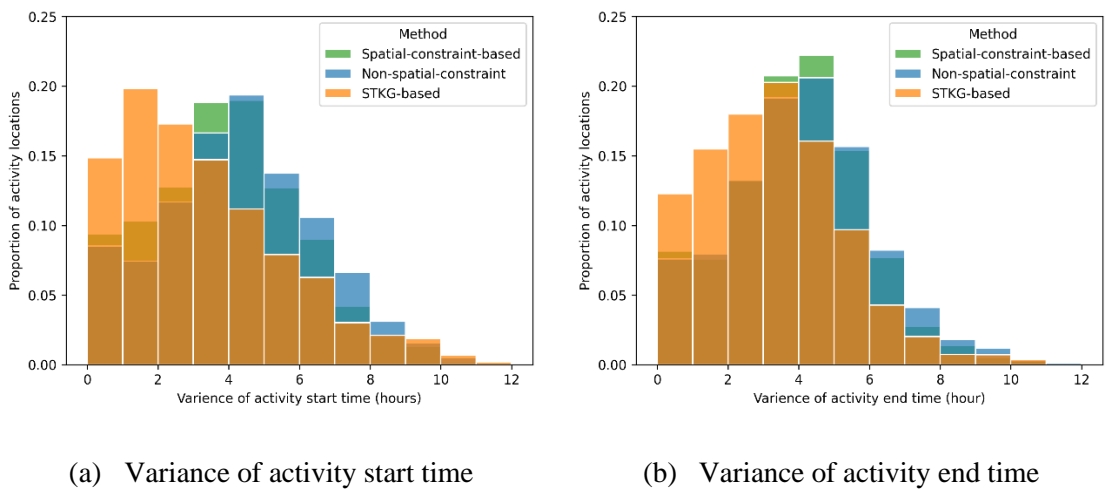

(a) Variance of activity start time          (b) Variance of activity end time

**Figure 8 Result of variance in start and end times of activities with the longest daytime stay**

Furthermore, **Figure 9** illustrates the joint distribution of the variance in start times of activities at locations with the longest daytime stay (potential workplaces) and the number of days these activities can be observed to occur. The observable days for work activities identified by the spatial-constraint-based method (8-12 days) are higher than those identified by the non-spatial-constraint method (4-8 days). The lack of spatial constraints in the non-spatial-constraint method leads to the identification of larger activity locations (as discussed earlier). As a result, daily work activities are confused with other activities, making it difficult to distinguish work activities for each day. However, the variance of activity start times for the spatial-constraint-based method is comparable to that of the non-spatial-constraint method (around 4 hours).

This indicates that the spatial-constraint-based method may prematurely interrupt or extend regular work activities.

In contrast, the joint distribution for the STKG-based method shows that high-density values are concentrated in the region with high observable days (13-15 days) and low variance of activity start time (around 2 hours). This further indicates that not only do the identified activity locations have strong regularity in activity start times, but such activities are also observed to occur stably and repeatedly.

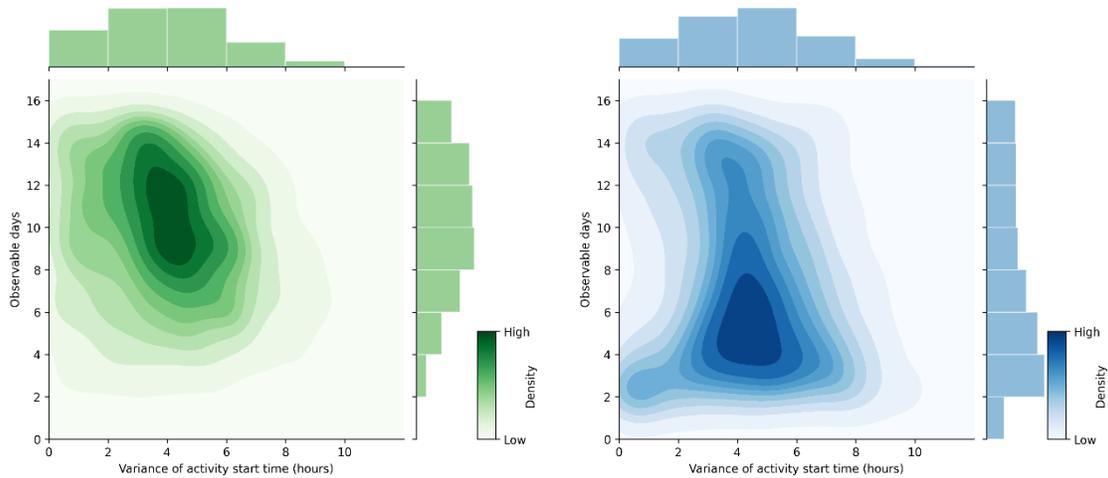

(a) Spatial-constraint-based method      (b) Non-spatial-constraint method

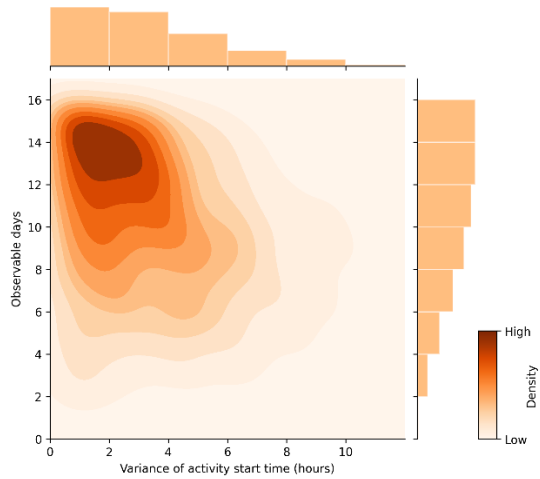

(c) STKG-based method

**Figure 9 Joint distribution of variance in start times of activities with the longest daytime stay and the number of days these activities are observable**

## 6.3 Impact on Activity Pattern Analysis

Th example of an individual is used to illustrate the impact of the identification of activity locations on activity pattern analysis. As depicted in **Figure 10**, the activity locations were identified by the STKG-based method and the two baseline methods. The time allocation of corresponding activities is also shown in **Figure 10**. This user's activity space is concentrated within an area of approximately 3 km × 3 km. The STKG-based method divides the activity space into two activity locations, with one overlapping grid cell and one non-adjacent grid cell. In contrast, the spatial-constraint-based method divides the activity space into three distinct activity locations under the spatial constraint, while the non-spatial-constraint method considers the activity space as a single activity location. Consequently, if the activity pattern analysis is based on the activity locations identified by the non-spatial-constraint method, this user's movements between different activity locations cannot be observed, indicating no mobility.

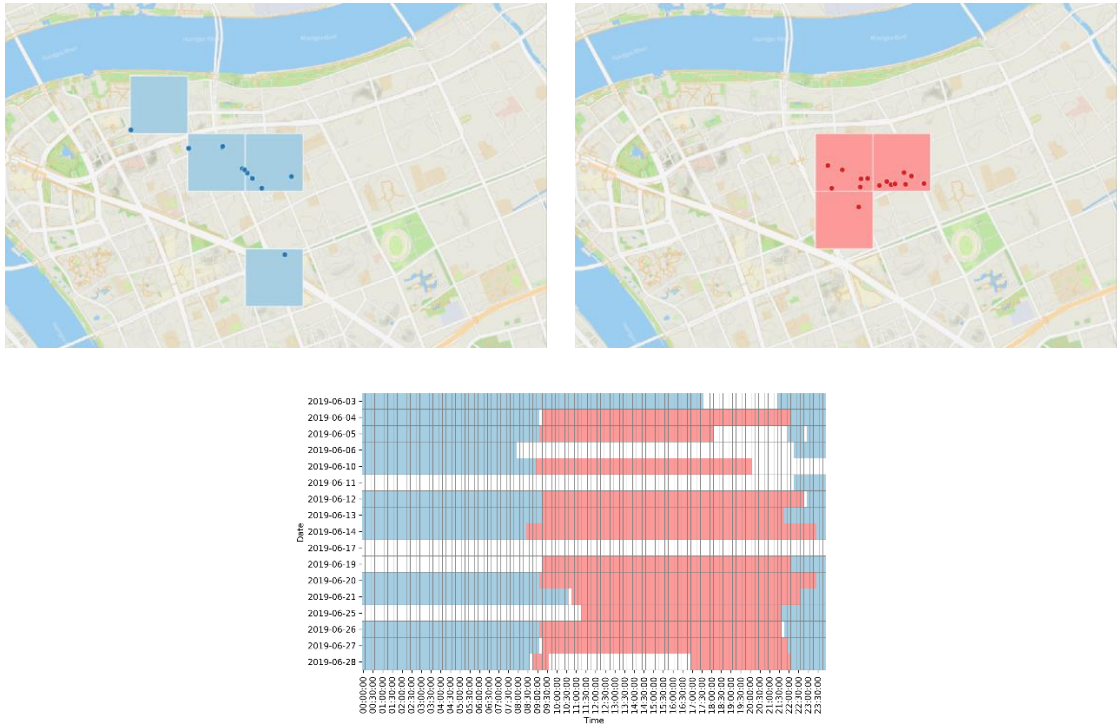

(a) STKG-based method

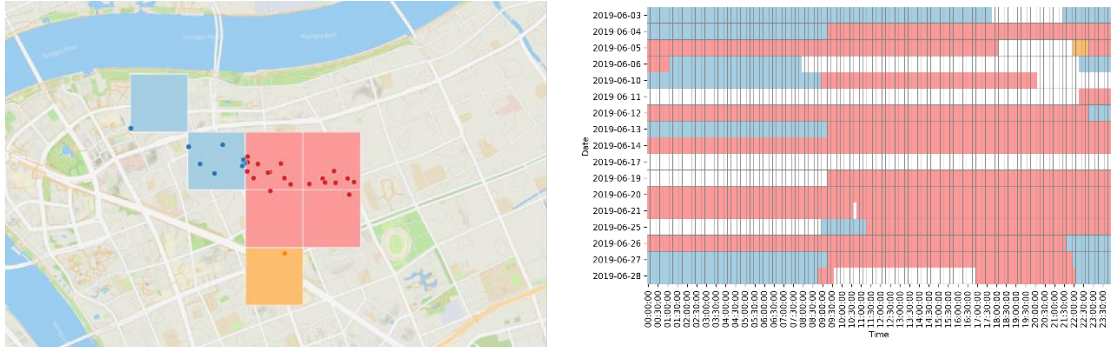

(b) Spatial-constraint-based method

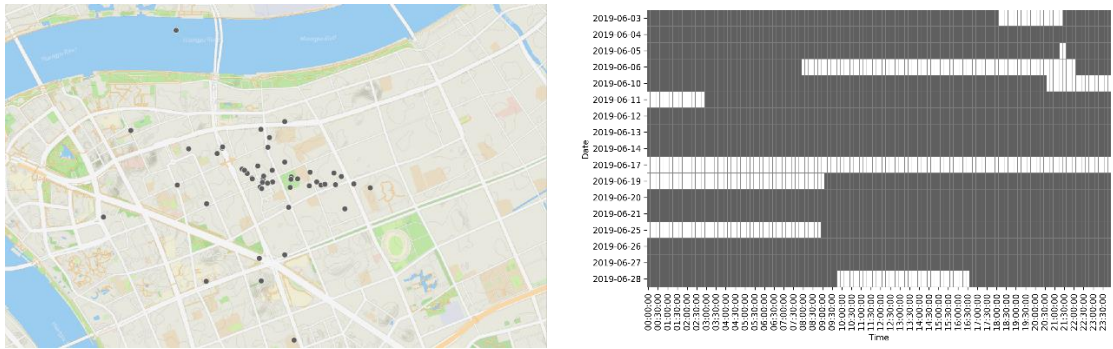

(c) Non-spatial-constraint method

**Figure 10 Activity locations and activity time allocation of an individual (each color represents a activity locations and its corresponding activity)**

Although the spatial distribution appears reasonable, the activity time allocation reveals some inconsistencies in the identification of activity locations by the spatial-constraint-based method. For spatial-constraint-based method, each grid is assumed to represent a single activity location, with the assumption that all stays within the grid are homogeneous. However, in the time allocation diagram based on this one-to-one correspondence (left side of **Figure 10(b)**), stays within the same activity location occurred at different times, such as during both the daytime and nighttime or across different days. As a result, while the spatial-constraint-based method successfully distinguishes locations that are close in proximity, the identified activity locations lack clear temporal differentiation.

In comparison, the two activity locations identified by STKG-based method are corresponding to daytime activities and nighttime activities, respectively. This provides potential information for home and out-of-home activities and distinguishes them both spatially and temporally, aiding subsequent activity pattern analysis. Although there are spatial overlaps, this is because the activity space is small, where the locational uncertainty of mobile phone data might shift stays to another activity location.

# 7 CONCLUSIONS

This study presents an STKG-based method for identifying individual activity locations from mobile phone data. The STKG is designed to model individual mobility patterns by incorporating both the spatial and temporal links between stays, which are subsequently transformed into a spatiotemporal graph. A modularity-optimization community detection algorithm is applied to detect stays with strong spatiotemporal connections, treating these as activity locations. A case study in Shanghai was conducted to validate the proposed method's performance and advantages. The experimental findings reveal that:

(1) Compared to the non-spatial-constraint method, even without explicitly imposing spatial constraints, the STKG-based method can limit an additional 45% of activity locations with the longest daytime stay within a reasonable spatial range.

(2) Activity locations identified by the STKG-based method exhibit lower variance in the start and end times of activities across different days, performing approximately 10% to 20% better than the two baseline methods. The number of observable days for frequent activities in the STKG-based identification results (13-15 days) is also higher than those in the spatial-constraint-based (8-12 days) and non-spatial-constraint methods (4-8 days), demonstrating the effectiveness of the STKG-based method in terms of temporal stability.

(3) The STKG-based method effectively distinguishes between locations that are geographically close but exhibit different temporal patterns, offering more precise insights for individual activity pattern analysis.

The challenge of validation is commonly faced by most human mobility studies based on mobile phone data (Huang et al., 2019). In the future, we hope to recruit volunteers to compare and validate their activity location identified based on passive mobile phone data with actively collected activity information, such as trip diaries. Another feasible approach is that the validation of the STKG-based method could be achieved at an aggregate level, which may facilitate easier access to validation datasets. In practice, our goal is to apply the proposed STKG-based method in transportation planning and policy-making to verify its effectiveness.

## ACKNOWLEDGMENTS

This study was supported by the Science and Technology Commission of Shanghai Municipal under Grant (22511104200).

24Zeng, J., Tang, J., 2023. Combining knowledge graph into metro passenger flow prediction: A split-attention relational graph convolutional network. Expert Systems With Applications 213, 118790.24